\pdfoutput=1

\documentclass[12pt]{iopart}

\usepackage{graphicx}

\usepackage{color}
\usepackage{float}

\begin{document}
\title{Radiation Pressure Acceleration of Thin foils with Circularly Polarized Laser Pulses}
\author{A.P.L Robinson$^1$, M.Zepf$^2$, S.Kar$^2$, R.G.Evans$^{1,3}$, C.Bellei$^3$}
\address{$^1$Central Laser Facility,STFC Rutherford-Appleton Laboratory, Chilton, OX11 0QX, UK}
\address{$^2$Department of Physics and Astronomy, Queen's University Belfast, BT7 1NN, UK}
\address{$^3$Plasma Physics Group, Blackett Laboratory, Imperial College London, SW7 2BZ, UK}
\date{\today}
\begin{abstract}
A new regime is described for Radiation Pressure Acceleration of a thin
foil by an intense laser  beam of above ~$10^{20}\mbox{Wcm}^{-2}$. Highly monoenergetic proton beams extending to GeV energies can be produced with very high efficiency using circularly polarized light.  The proton beams have a very small divergence angle (less than 4 degrees). This new method allows the construction of ultra-compact proton and ion accelerators with ultra-short particle bursts.
\end{abstract}

\maketitle
%
\section{Introduction}
Since the pioneering work of Clark et al. \cite{exp2} and the studies done with the Nova PetaWatt laser \cite{exp1,hatchett1}, there has been considerable interest in the generation of multi-MeV proton beams from thin foil targets irradiated by ultraintense lasers \cite{esirkepov1,mora1,schwoerer1,hegelich3,marcoscience,exp20,exp9,exp6}.   
This interest is driven by some outstanding qualities of these beams ­ their very low emittance \cite{exp20} , short burst duration and the very large accelerating field ($ >10^{12} \mbox{Vm}^{-1}$). The large accelerating field allows ions to be acclerated to MeV type energies over micron length scales and hence allows the construction of ultra compact accelerators. Potential applications that might benefit from an accelerator with such properties are Fast Ignition Inerital Confinement Fusion (ICF) \cite{tabak,rothprotonfi,temporal} and radiotherapy for the treatment of tumours and probing laser-plasma interactions \cite{marcoprobe}.
Currently the extremely low emittance is achieved by a combination of large divergence with a small source size and short burst duration with a 100\% energy spread. Any novel scheme that can reduce the divergence and produce monoenergetic beams would therefore present a very substantial breakthrough.
In experiments to date the acceleration normally attributed to Target Normal Sheath Acceleration (TNSA) \cite{mora1,bychenkov,wilks1,sentoku1,kemp3}. In TNSA the acceleration of the initially cold ions is due to the strong fields set up by a sheath of laser accelerated relativistic electrons at the rear of the foil. The ion energy is dominated by the temperature and density of the hot electron population and hence scales as $(I\lambda^2)^{1/2}$.

Another, in principle very efficient means of using lasers to accelerate foils to high velocities has been discussed in the literature for many years \cite{marx} - Radiation Pressure Acceleration (RPA). Hereby the momentum of the laser is imparted directly to the object to be accelerated. Simple analytic models based on momentum conservation indicate that the final energy should simply scale as $\propto (I\tau/\sigma)^{\alpha}$ (where $I$ is the intensity,$\tau$ the pulse duration, and $\sigma$ is the areal mass of the foil). The exponent, $\alpha$ is equal to 2 for $v_{final} \ll c$, and $\alpha \rightarrow 1/3$ in the ultrarelativistic limit.  In terms of simple numbers, any laser capable of  $>10^7 \mbox{Jcm}^{-2}$ should therefore be able to accelerate sub-micron foils to a monoenergetic beam with energies in excess of 1 MeV and should provide a much more favourable intensity scaling than TNSA (which scales $\propto I^{1/2}$) in most cases.  However, this has never been demonstrated for realistic parameters above the plasma formation threshold.

While high power lasers that can deliver $>10^7 \mbox{Jcm}^{-2}$ in a variety of pulse durations and wavelengths have existed for decades, there is no experimental evidence of RPA actually being the dominant process anywhere in this wide parameter range. For example for lasers with $>10^{15} \mbox{Wcm}^{-2}$ and $10\mbox{ns}$ duration, foil acceleration is ablatively driven \cite{lindl} and for CPA lasers with $>10^{19} \mbox{Wcm}^{-2}$ and $<1\mbox{ps}$ TNSA is the dominant particle acceleration mechanism \cite{hatchett1}. The question is therefore whether there is any scheme for realistic laser parameters in which RPA can be expected to dominate. Recently this area has been invigorated by the numerical demonstration of RPA by Esirkepov et al. \cite{esirkepov2} at $I>10^{23}\mbox{Wcm}^{-2}$ using linear polarisation. Unfortunately this high intensity will preclude the possibility of testing this theory for a number of years and making a practically working system with reasonable repetition rates an even more distant prospect.

In this paper we identify for the first time a realistic RPA scheme for current laboratory lasers - based on the use of circular polarization to effectively couple the laser to the entire foil. Simulations using 1D and 2D PIC simulations predict a step change in the performance of laser based proton and ion accelerators -making low divergence, monoenergetic GeV beams a realistic prospect. Comparisons with the analytical model suggest that we have indeed identified a scheme in which RPA is the dominant interaction.

This paper has four main parts: A discussion of the basic mechanism of our new scheme and comparisons to the scaling predicted by the analytical RPA model, a discussion of the ion energy scaling, some of the practical limits (failure modes) and a discussion of the 2D calculations. The majority of the simulations has been carried out using the  electromagnetic (EM)  Particle in Cell (PIC) method with one spatial and three momentum dimensions (1D3P). Higher dimensional effects have been addressed by 2D3P calculations using the OSIRIS code \cite{osiris}.

\section{Theory}
The theoretical starting point for this work is the 1D model of radiation pressure acceleration\cite{marx}. The 1D relativistic equation of motion for an illuminated foil is derived by first noting that the intensity of the incident radiation in the instantaneous rest (primed) frame of the foil is $I^{\prime}/I  = (1 - v/c)/(1 + v/c)$ where $v$ is the foil velocity.  Since the 3-force is colinear with the foil velocity, $f^{\prime}_x = f_x$.  One therefore obtains the following equation of motion:

\begin{equation}
\label{eom}
\frac{dp}{dt} = \frac{2I}{c}\frac{\sqrt{p^2+\sigma^2c^2}-p}{\sqrt{p^2+\sigma^2c^2}+p},
\end{equation}

where $p$ is the areal momentum of the foil, and $\sigma$ is the areal mass of the foil.  The most elegant way to integrate Eq.\ref{eom} is to seek a solution of the form,

\begin{equation}
\label{subform}
p = \sigma{c}(\sinh(\psi) - 1/4\sinh(\psi)).
\end{equation}

One then immediately finds that:
 
\begin{equation}
\label{sol2}
\psi = \frac{1}{3}\sinh^{-1}\left[\frac{6It}{m_in_ilc^2}+2\right],
\end{equation}

where we have written the foil areal mass as $\sigma = m_in_il$ ($m_i$ is the ion mass,$n_i$ the ion density, and $l$ is the foil thickness).  The momentum of an individual ion is then given by $p_i = m_ic(\sinh(\psi) - 1/4\sinh(\psi))$.

If one illuminates at a constant intensity of, $I = 2 \times 10^{21}\mbox{Wcm}^{-2}$, for 400fs, and the foil has an areal mass of $5 \times 10^{-5}\mbox{kgm}^{-2}$ (a 50nm thick carbon foil at 1000kgm$^{-3}$), then one finds that the resultant energy per carbon ion is 5.5 GeV.  This is a considerable energy by the standards of most experimental observations, indeed it is well in excess of the 1--2 GeV required for certain medical applications, and the energy spectrum is instrinsically monoenergetic.  

Substantial caution, however, needs to be applied, since this theory makes the assumption that the laser momentum is efficiently coupled directly to the ion population.  The laser primarily interacts with the plasma electrons, and usually the heating of the electrons leads to sheath acceleration dominating. This prevents the entire foil from being uniformly accelerated. For any RPA scheme it is essential that the electric field is dominated by a single positive spike (front and back sheaths must be small in comparison).  This magnitude of this spike must be approximately $E_x = F_{rad}/en_{e,0}l$ for RPA to dominate (this also implies that the electrons are in force balance), otherwise equation \ref{eom} will not model the foil motion.  If the size of the spike is comparable to the foil thickness then, from Gauss' Law, one finds that the electron and ion densities on each side of the spike must satisfy $2I\epsilon_0/e^2n_{e,0}cl^2 \approx |Zn_i-n_e|$.  For $I = 10^{21}\mbox{Wcm}^{-2}$ and $l=100$nm one concludes that about half the electrons from one side of the foil need to be moved into the other, and this must be sustained in a quasi-static fashion.

\section{Simulation Results}
The challenge is then to find an interaction regime in which these conditions are met, and competing mechanisms (e.g TNSA) are suppressed.  Our results show that this can be achieved by illuminating a solid target with ultra high contrast pulses with circular polarization. In the case of circular polarization the electron dynamics are substantially different. In particular the ponderomotive pressure no longer oscillates through zero twice per optical cycle. Instead the ponderomotive pressure becomes slowly varying and follows the shape of the intensity envelope.  This allows the strong, sustained, expulsion of electrons at lower intensites than with linear polarization which results in weaker expulsion at a given intensity.  Circular polarization also reduces the fast electron energy thus suppressing the sheath fields. The only alternative appears to be to exploit the strong intensity scaling of RPA vs TNSA and increase the intensity until RPA dominates the interaction. Unfortunately, this only occurs for laser parameters which are orders of magnitude above the current state of the art \cite{esirkepov2}.

In numerical simulations it is observed that, initially a strong expulsion of electrons occurs in a very thin region.  As the spike in the electric field accelerates ions out of this thin region and the critical surface retreats along with the ions the laser begins penetrate beyond the initial target surface.  The ions behind this initial layer are then accelerated similarly.  This particular process appears to be very similar to that described by Macchi et al. \cite{macchi1}.  However, as the foil is very thin, eventually the entire ion population is accelerated in this way and has achieved a net positive momentum.  Then the ions that originated from the front surface are eventually overtaken, and re-accelerated.  This repeats in cyclic fashion. 

\begin{figure}[ht]
\begin{center}
\includegraphics[width=\textwidth]{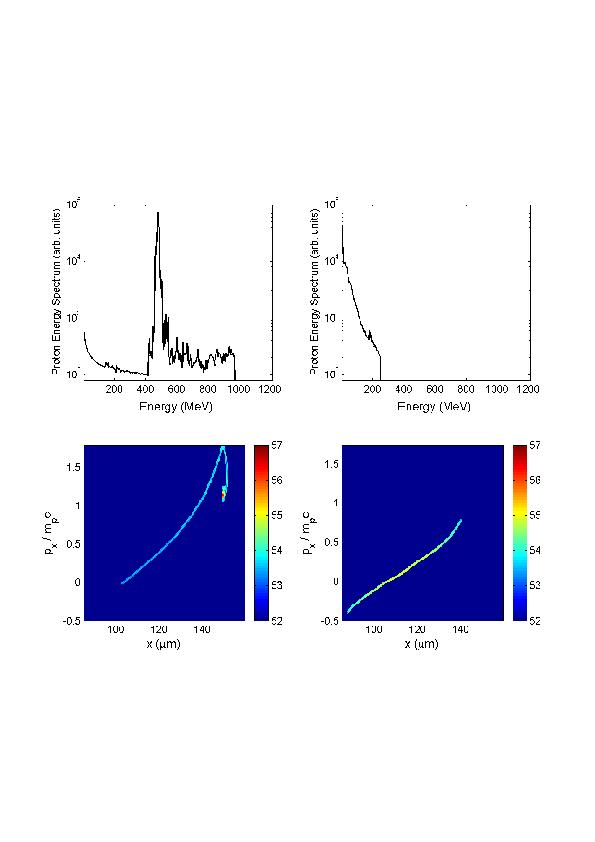}
\caption{\label{fig:fig1}(Top Left) Proton energy spectrum after 400fs in the standard run. (Bottom Left) $\log_{10} p_x-x$ proton phase space at 400fs in standard run.(Top Right)Proton energy spectrum for identical parameters but with linear polarization. (Bottom Right) $\log_{10} p_x-x$ proton phase space at 400fs in linear polarization run.  In the case of circular polarization the foil is accelerated as a whole, leading to a narrow energy spectrum, whereas in the case of linear polarization the foil violently decompresses and produces a broad energy spectrum.}
\end{center}
\end{figure}

It should be stressed that Macchi et al. did not consider or demonstrate the possbility of an entire foil being accelerated in the RPA regime.  Macchi et al. concluded that their scheme could produce 'moderate' energy ions, whereas this work demonstrates that this can easily be extended to produce high energy ions.  We should also address the matter of the statement in the later work of Esirkepov et al. \cite{esirkepovparametric} that a `transition' into the RPA regime begins around $5 \times 10^{21} \mbox{Wcm}^{-2}$.  Exactly how a smooth transition occurs was not made clear.  In the RPA mechanism there is no dependence on the ion charge or laser wavelength, for example, whereas there is a strong dependence the TNSA mechanism.  Nor was it demonstrated that strongly monoenergetic spectra could be obtained early in this transition.  In contrast to that work, what we are demonstrating here is that a {\em complete} switch from the TNSA to the RPA regime can be obtained at intensities around $10^{21}\mbox{Wcm}^{-2}$.  This is illustrated clearly by figure \ref{fig:fig1}.


The specifics of the 1D EM PIC simulations are now reported.  This considered a 150nm foil consisting of protons at density of $8 \times 10^{28}\mbox{m}^{-3}$ with a corresponding electron density to give initial charge neutrality with an initial electron and ion temperature of 10keV. The front surface is initially located at 105$\mu$m on the grid.  A circularly polarised EM wave with $\sin^4(t)$ with a Full Width Half Maximum (FWHM) pulse duration of 64fs and the peak intensity is $2 \times 10^{21}\mbox{Wcm}^{-2}$ is incident on the foil.  This constitutes our standard calculation, which we carry out up to 400fs.  As anticipated the foil is accelerated as a whole and the spectrum shows that almost the entire ion population is concentrated in a monoenergetic spectral peak at 485 MeV (Fig. \ref{fig:fig1}(Top Left)). 

\begin{figure}[ht]
\begin{center}
\includegraphics[width=\textwidth]{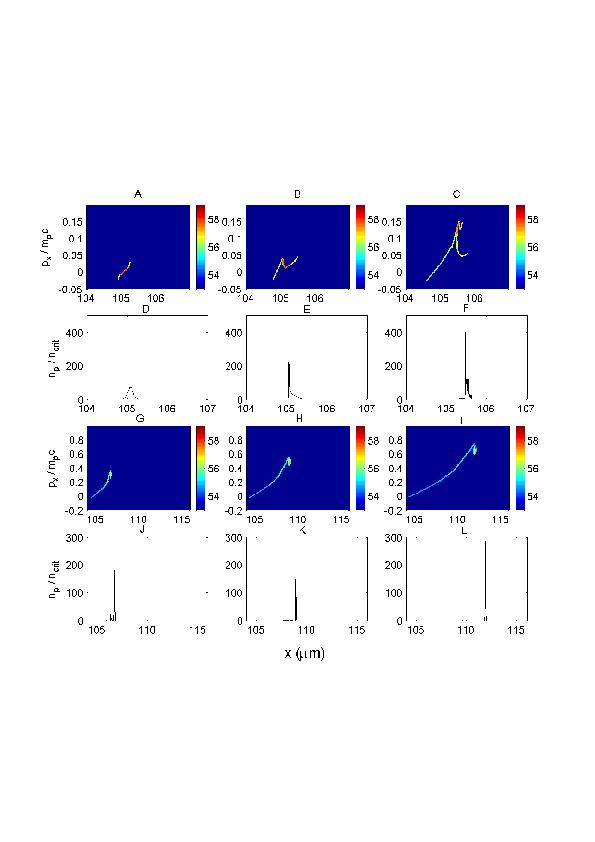}
\caption{\label{fig:fig1b}(A--C, G--I)Proton $\log_{10} p_x-x$ phase space plots in the standard run. (D--F, J--L)Proton density ($n_p/n_{crit})$) profiles in the standard run. Times are: A/D = 20fs, B/E = 40fs, C/F = 60fs, G/J = 80fs, H/K = 100fs, I/L = 120fs.  Note the changes in scale.  It can be seen that, after 60fs, the foil moves as a whole with only a very small amount of material being left as trailng mass. }
\end{center}
\end{figure}

The aforementioned evolution of the proton phase space is illustrated by figure \ref{fig:fig1b}.  The corresponding proton density profiles are also shown.  This shows how the ions are initially accelerated at the front surface and are driven through the target.  It is during the time, from 20-60fs, that the phase space evolution has many similarities to the process described by Macchi et al..  It is from 60fs onwards that the cyclic acceleration of the foil mass occurs, i.e. after the ions from the front surface have been pushed through the rear surface.  At these later times the phase space shows a distinct `head', where the majority of the ion mass is concentrated, and a `tail' which contains a very low amount of trailing mass.

\begin{figure}[ht]
\begin{center}
\includegraphics[width=\textwidth]{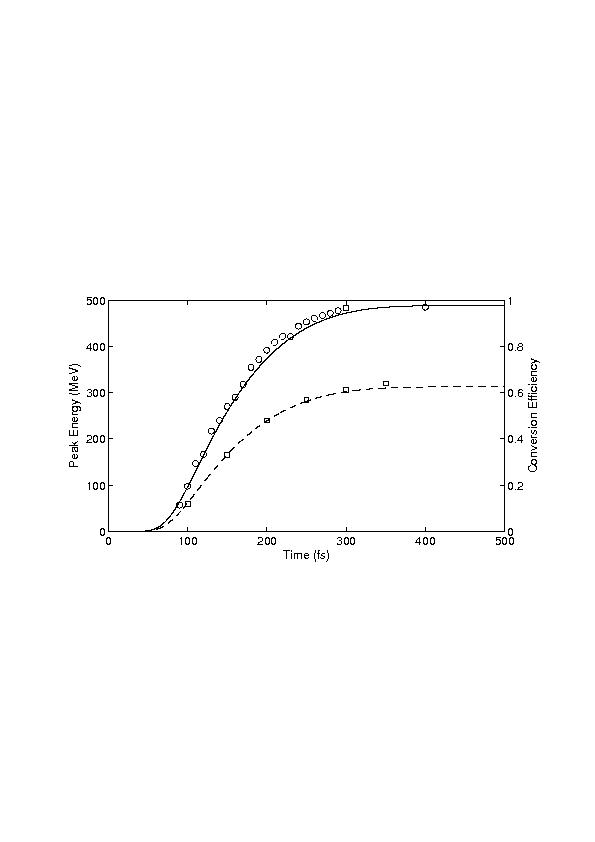}
\caption{\label{fig:fig2}Comparison of semi-analytic model (solid line) to peak proton energy in the standard run (circles).  Conversion efficiency also shown for both the model (dashed line) and standard run (squares).}
\end{center}
\end{figure}

The results of the simulation are in excellent agreement with a semi-analytical RPA model as can be seen in figure \ref{fig:fig2}, where the prediction of a semi-analytical model for the energy of the spectral peak (i.e. modal energy) and the conversion efficiency (reaching $>$60\% ) is compared to the simulation results.  The semi-analytic model is based on integrating Eq. \ref{eom} and an equation which specifies the intensity profile (i.e.$I = I(t)$).  The position of the foil, $x_{foil}(t)$, must also be be tracked so one has two first order differential equations and a pulse specification equation which can be integrated numerically very quickly. The peak energy in the simulation is always slightly higher than that predicted by the semi-analytic model.  This is because the foil loses some mass in the form of `trailing mass' that gets left behind as the foil is accelerated.  

It is also interesting to assess factors that do not play any role in the analytic model.  Firstly there is the matter of wavelength.  This is absent from the analytic model (as seen in Eq.\ref{eom}), and this contrasts greatly with TNSA in which ion energy scaling depends on $(I\lambda^2)^{1/2}$ \cite{wilks1}.  On repeating the standard run with 0.5$\mu$m light, it is  found that the results differ less than 10\%, however this must become important in the limit where the wavelength becomes so short that the foil is transparent.  Secondly there is the matter of the ion charge, or the charge to mass ratio of the ions, which is also absent from the analytic model, but is not in TNSA \cite{mora1}.  When the standard run is repeated for particles with $A=1$ and $Z=2$ the results also differ by less than 10\%. Thus it has been shown that the radiation pressure model is an excellent explanation of these simulation results both in terms of the peak proton energy evolution with time and the fact that this mechanism has only a very weak dependence on wavelength or ion charge.  

It is instructive  to contrast these results with the results obtained using linear polarization in otherwise identical conditions. The constant expulsion of electrons is not achieved and this acceleration mechanism fails giving way to TNSA (Fig. \ref{fig:fig1}(Top Right)), in agreement with experimental evidence \cite{mckenna_contrast}.  The phase space plots in figure \ref{fig:fig1}(Bottom Left and Right) also show that the acceleration is completely different in the two cases.  We should re-iterate that this shows that the change in laser polarization results in a {\em} complete switch between two completely different acceleration regimes. 


\section{Discussion}
\begin{figure}[ht]
\begin{center}
\includegraphics[width=\textwidth]{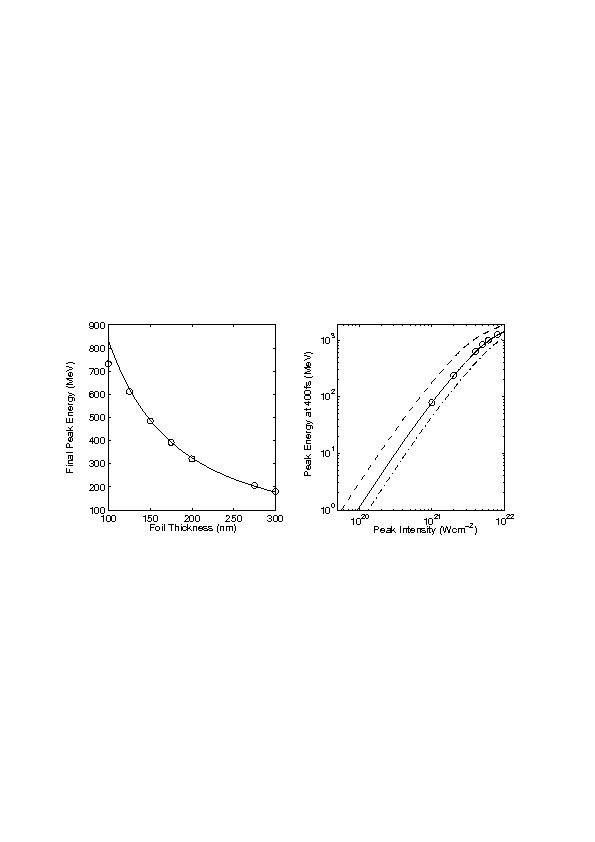}
\caption{\label{fig:fig3} Peak proton energy against foil thickness (Left) and peak intensity (Right) in both PIC simulations (circles) and semi-analytic model (lines). Dashed:150nm, solid:250nm, dash-dot:350nm foil thickness in (Right).}
\end{center}
\end{figure}

The semi-analytical model allows the dependence on the final peak proton energy on foil thickness d and intensity to be estimated very accurately.  These scale as $\propto d^{-1}$ and $\propto I$ (for non-relativistic foil velocities) respectively.  This was shown to be the case by carrying out a set of 1D PIC runs over a wide range of parameters, the results of which are shown in (Fig. \ref{fig:fig3}).  As in the case of the standard run, the agreement with the semi-analytic model is excellent.  Monoenergetic spectra are still obtained in each case, although the relative energy spread is larger the lower the final proton energy.  The highest proton energies are thus achieved for thin foils at high irradiance with 1GeV protons predicted for intensities $~3 \times 10^{21}\mbox{Wcm}^{-2}$ and d=150nm. 


To assess how robust this scheme was, several potential 'real-life' failure modes were investigated. Clearly, the foil must be thick enough and have a sufficiently high density to avoid becoming transparent.  In the case of the standard run (1$\mu$m wavelength) this begins to occur below 100nm (see figure \ref{fig:fig3}(Left)), suggesting that the acceleration of ultrathin foils at very high intensities is only possible using longer wavelengths.  Complete failure for the standard run parameters was found to occur at 20nm foil thickness.  The effects of the pulse being elliptically, as opposed to perfectly circularly, polarized were also investigated. It was found that a deviation from circular to ellipticities of 1.5:1 to 2:1 had little significant effect.  

A critical failure mode that is relevant to any RPA scheme is also highlighted for targets consisting of multiple ion-species. Since the electric field will be determined by the electron density $n_e = Z_1n_{i,1} + Z_2n_{i,2}$, and the equation of motion for each ion species will be given by:

\begin{equation}
\label{exmix}
\frac{dp_{i,k}}{dt} = \frac{Z_kn_{i,k}\frac{2I}{c}\frac{1-v/c}{1+v/c}}{(Z_1n_{i,1} + Z_2n_{i,2})}.
\end{equation}

This scenario was examined by comparing simulations identical to the standard run, except that the protons were replaced by a mixture of protons and C$^{6+}$ each at a density of $4 \times 10^{28}\mbox{m}^{-3}$. The acceleration of the C$^{6+}$ ions was not significantly affected and reached a final energy of 280MeV compared to 339 MeV predicted by the semi-analytic model. By contrast the protons were badly affected and only reached energies of 20-40 MeV compared to 485 MeV achieved in the standard run. Although the species separated, the foil was not destroyed in this instance. However, the protons can only ever be accelerated to the extent that they are co-moving with the heavy ions, hence achieve a greatly reduced energy.  Therefore accelerating protons using this scheme presents technical challenges, but the acceleration of heavy ions is easier as proton contamination is uncritical, which stands in stark contrast to the TNSA mechanism.

\begin{figure}[ht]
\begin{center}
\includegraphics[width = \textwidth]{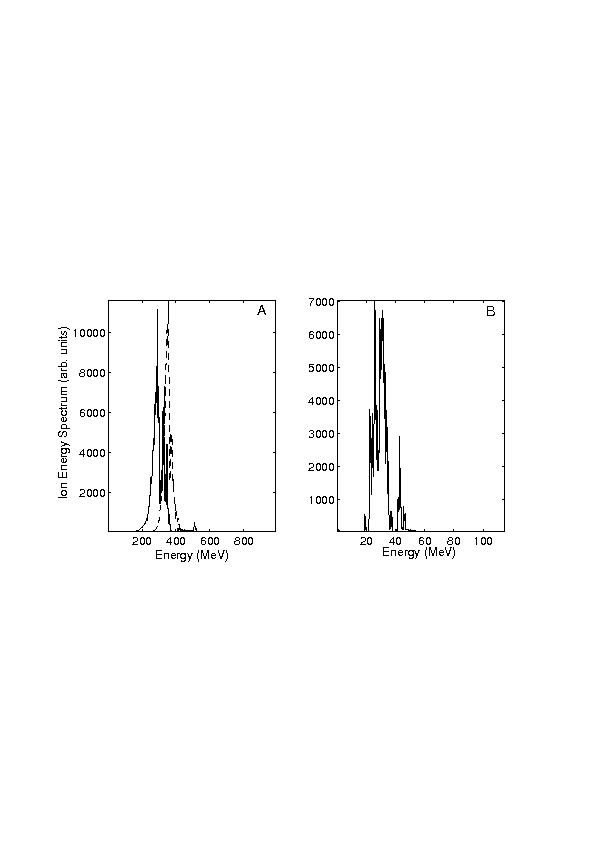}
\caption{\label{fig:fig3b}(Left)C$^{6+}$ energy spectra in 1:1 C$^{6+}$:H$^{+}$ simulations (solid) and pure C$^{6+}$ simulation (dashed).(Right)Proton energy spectrum in 1:1 C$^{6+}$:H$^{+}$ simulation.}
\end{center}
\end{figure}

This scenario was examined by comparing simulations identical to the standard run, the results of which are shown in fig. \ref{fig:fig3b}.  The run is identical to the standard run, except that the protons have been replaced by a mixture of protons and C$^{6+}$ each at a density of $4 \times 10^{28}\mbox{m}^{-3}$.  The results for a simulation where the protons are not present is also shown.  The energy spectra are taken from the simulation at 400fs in all three cases.  The energy spectrum of the C$^{6+}$ ions in fig. \ref{fig:fig3b} show that the C$^{6+}$ ions are not badly affected by this failure mode.  Note that the semi-analytic model predicts an energy of 339MeV for the C$^{6+}$ only case.  The protons are undesirable however as a lower peak energy is achieved.  The protons are badly affected by this failure mode however, as shown by the proton energy spectrum in fig. \ref{fig:fig3b} in which proton energies between 20-40 MeV are achieved which should be compared to the 485MeV achieved in the standard run.  In the simulation it is observed that the protons move ahead and separate from the C$^{6+}$ ions.  The accelerating electric field is much lower in the proton-only region.  The C$^{6+}$ ions subsequently overtake the protons, and the protons are accelerated again.  The result is that the protons are only ever accelerated to such an extent that they can move at approximately the same speed as the carbon ions, and thus only low proton energies are achieved.
Therefore accelerating protons using this scheme presents technical challenges, but the acceleration of heavy ions is easier since proton contamination is unimportant.  This stands in stark contrast to the TNSA mechanism in which the presence of a light ion population strongly inhibits the acceleration of heavy ions.

Since the foils must be very thin for efficient RPA to take place, pulse contrast becomes a critical issue. Recent experiments \cite{neely1} have demonstrated that ultrathin foils as thin as 50nm can remain intact until the peak of a 50fs, $10^{19}\mbox{Wcm}^{-2}$ pulse arrives. This demonstrates that sufficiently high contrast can be achieved, although plasma mirrors might be required at the highest intensities \cite {dromeyPM}. However, there will inevitably be some decompression as a result of the rising edge of even ultra-high contrast pulses for such thin foils. To test that the scheme is not dependent on the precise density profile, simulations were performed on foils that were decompressed by a factor of two in density to simulate the effect of the rising edge of the pulse. This also showed very little effect on the overall acceleration of the foil.  

\section{Two Dimensional Simulations}
Finally, to assess the impact of higher dimensional effects,  2D3P OSIRIS simulations were performed.  These simulations used a grid of 4000 cells along the laser axis and 8000 cells transversely (~16 $\mu m$ $\times$ 32$\mu m$).  The foil was 100~nm thick consisting of electrons and protons (400 particles per cell at a density of 100~$n_{crit}$,  initial electron temperature of 3keV).  The normally incident laser pulse was circularly polarised with a peak $a_0$ of 34 and was approximately Gaussian with 31fs FWHM duration.  The transverse spatial distribution is a fourth order supergaussian with a 1/e width of ~13$\mu m$ - corresponding to focusing top-hat laser beam outside the Rayleigh range.  This shape contributes greatly to the flat acceleration profile, a normal Gaussian gives a very curved shape to the foil. 

\begin{figure}[ht]
\begin{center}
\includegraphics[width = \textwidth]{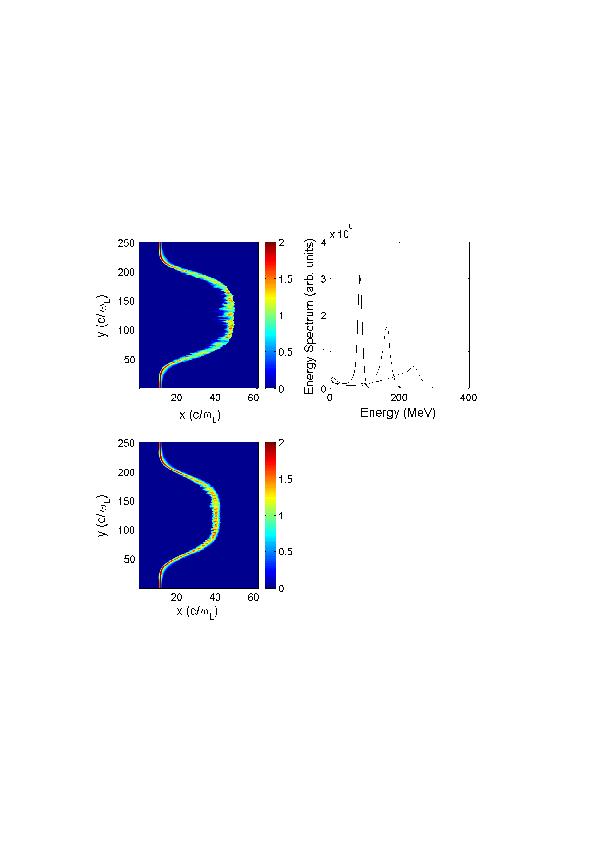}
\caption{\label{fig:fig6}Results from 2D OSIRIS Simulation:(Top Left)Proton density ($log_10(n_p/n_{crit})$) at 64fs. (Bottom Left)Proton density ($log_10(n_p/n_{crit})$) at 64fs in simulation with zero initial electron temperature. (Top Right)Proton energy spectra at 53 (solid), 64 (dashed), 75 (dash-dot) fs.}
\end{center}
\end{figure}

Figure \ref{fig:fig6} shows both a plot of the ion density at 64~fs, and the energy spectrum at different times.  The divergence angle of the protons (within $y=$10-20$\mu$m) is less than 4 degrees.  In 1D, taking the on-axis intensity, the semi-analytic model predicts peak energy of 260~MeV, compared to a peak energy of 240~MeV in the 2D calculation.  This shows that 2D effects do not substantially affect the acceleration process.  As one would expect, the foil shows some transverse Rayleigh Taylor like instability which begins to result in spectral broadening at late times. It should be emphasized that the growth rate and hence the scale of spectral broadening is overestimated by the current simulations.  It was found that these effects decrease substantially as the number of particles per cell is increased and initial temperature is decreased.  The effect of setting the initial electron temperature to zero is also shown in figure \ref{fig:fig6}.  A further increase in the number of particles per cell and a reduction of cell size (allowing a lower initial temperature) is expected to reduce these effects further. However, these simulations represent the limit of our computational capacity.  

\section{Summary}
In summary, it has been demonstrated that laser based ion acceleration based on radiation pressure is viable at intensities around $10^{20}-10^{21}~\mbox{Wcm}^{-2}$ using high contrast, circularly polarized pulses.  This produces highly energetic, monoenergetic ions very efficiently.    The mechanism has been identified as a radiation pressure acceleration on the basis of a semi-analytic model.  The semi-analytic model agrees extremely well with the simulation results and thus accurately predicts the energy scaling. Radiation pressure acceleration of ions at intensities achievable with the latest laser technology represents a step change in the quality and performance of laser-ion acceleration.

\section{Acknowledgements}
The authors are grateful for the use of computing resources provided by STFC's e-Science facility.  M.Zepf is supported by the Royal Society.  

\bibliographystyle{unsrt}
\bibliography{references}

\begin{thebibliography}{10}

\bibitem{exp2}
E.L.Clark et~al.
\newblock Energetic heavy-ion and proton generation from ultraintense
  laser-plasma interactions with solids.
\newblock {\em Phys. Rev. Lett.}, 85(8), 2000.

\bibitem{exp1}
R.A.Snavely et~al.
\newblock Intense high energy proton beams from petawatt-laser irradiation of
  solids.
\newblock {\em Phys.Rev.Lett.}, 85(14):2945, 2000.

\bibitem{hatchett1}
S.Hatchett et~al.
\newblock {\em Phys.Plasmas}, 5:2076, 2000.

\bibitem{esirkepov1}
T.Z.Esirkepov et~al.
\newblock Proposed double-layer target for the generation of high-quality
  laser-accelerated ion beams.
\newblock {\em Phys.Rev.Lett.}, 89:175003, 2002.

\bibitem{mora1}
P.Mora.
\newblock Plasma expansion into vacuum.
\newblock {\em Phys. Rev. Lett.}, 90:185002--1 -- 185002--4, 2003.

\bibitem{schwoerer1}
H.Schwoerer et~al.
\newblock Laser-plasma acceleration of quasi-monoenergetic proton from
  microstructured targets.
\newblock {\em Nature}, 439:445, 2006.

\bibitem{hegelich3}
M.Hegelich et~al.
\newblock Laser acceleration of quasi-monoenergetic mev ion beams.
\newblock {\em Nature}, 439:441, 2006.

\bibitem{marcoscience}
T.Toncian et~al.
\newblock {\em Science}, 312:410, 2006.

\bibitem{exp20}
T.Cowan et~al.
\newblock Ultralow emittance, multi-mev proton beams from a laser
  virtual-cathode plasma accelerator.
\newblock {\em Phys.Rev.Lett.}, 92:204801--1, 2004.

\bibitem{exp9}
M.Borghesi et~al.
\newblock Multi-mev proton source investigations in ultra-intense laser-foil
  interactions.
\newblock {\em Phys.Rev.Lett.}, 92:05503--1 -- 05503--4, 2004.

\bibitem{exp6}
A.J.Mackinnon et~al.
\newblock Enhancement of proton acceleration by hot electron recirculation in
  thin foils irradiated by ultraintense laser pulses.
\newblock {\em Phys.Rev.Lett.}, 88(22), 2002.

\bibitem{tabak}
M.Tabak et~al.
\newblock Ignition and high gain with ultrapowerful lasers.
\newblock {\em Phys.Plasmas}, 1(5), 1994.

\bibitem{rothprotonfi}
M.Roth et~al.
\newblock {\em Phys.Rev.Lett.}, 86:436, 2001.

\bibitem{temporal}
M.Temporal et~al.
\newblock Numerical study of fast ignition of ablatively imploded
  deuterium-tritium fusion capsules by ultra-intense proton beams.
\newblock {\em Phys.Plasmas}, 9:3098, 2002.

\bibitem{marcoprobe}
M.Borghesi et~al.
\newblock Plasma ion evolution in the wake of a high-intensity ultrashort laser
  pulse.
\newblock {\em Phys.Rev.Lett.}, 94:195003, 2005.

\bibitem{bychenkov}
V.Yu.Bychenkov.
\newblock Ion acceleration in expanding multispecies plasmas.
\newblock {\em Phys.Plasmas}, 11:3242, 2004.

\bibitem{wilks1}
S.C.Wilks et~al.
\newblock Energetic proton generation in ultraintense laser-solid interactions.
\newblock {\em Phys.Plasmas}, 8:542, 2001.

\bibitem{sentoku1}
Y.Sentoku et~al.
\newblock High energy proton acceleration in interaction of a short laser pulse
  with dense plasma target.
\newblock {\em Phys.Plasmas}, 10:2009, 2003.

\bibitem{kemp3}
A.J.Kemp and H.Ruhl.
\newblock Multispecies ion acceleration off laser-irradiated water droplets.
\newblock {\em Phys.Plasmas}, 12:033105, 2005.

\bibitem{marx}
G.Marx.
\newblock Interstellar vehicle propelled by terrestrial laser beam.
\newblock {\em Nature}, 211:22, 1966.

\bibitem{lindl}
J.Lindl et~al.
\newblock The physics basis for ignition using indirect-drive targets on the
  national ignition facility.
\newblock {\em Phys.Plasmas}, 11:339, 2004.

\bibitem{esirkepov2}
T.Esirkepov et~al.
\newblock Highly efficient relativistic-ion generation in the laser-piston
  regime.
\newblock {\em Phys.Rev.Lett.}, 92:175003--1, 2004.

\bibitem{osiris}
S.Lee.
\newblock {\em Phys.Rev.E}, 61(1074), 1988.

\bibitem{macchi1}
A.Macchi et~al.
\newblock Laser acceleration of ion bunches at the front surface of overdense
  plasmas.
\newblock {\em Phys.Rev.Lett.}, 94:165003, 2005.

\bibitem{esirkepovparametric}
T.Esirkepov et~al.
\newblock {\em Phys.Rev.Lett.}, 96:105001, 2006.

\bibitem{mckenna_contrast}
P.McKenna et~al.
\newblock {\em Phil.Trans.R.Soc.A}, 364:711, 2006.

\bibitem{neely1}
D.Neely et~al.
\newblock {\em Appl.Phys.Lett.}, 89:021502, 2006.

\bibitem{dromeyPM}
B.Dromey et~al.
\newblock {\em Rev.Sci.Instrum.}, 75:645, 2004.

\end{thebibliography}

\end{document}